\def\aa{{A\&A}}

\def\aj{{AJ}}
\def\annrev{{ARA\&A}}
\def\apj{{ApJ}}
\def\apjs{{ApJS}}

\def\mnras{{MNRAS}}
\def\nat{{Nature}}
\def\pasp{{PASP}}

\documentclass[cup5b]{caps}
\usepackage{graphicx}
\usepackage{amssymb}
\usepackage{ociwsymp1}
\HeadText{P. S. Osmer}

\begin{document}

\pagenumbering{arabic}

\author[]{PATRICK S. OSMER\\Department of Astronomy, The Ohio State University}

\chapter{The Evolution of Quasars}

\begin{abstract}

This article reviews and discusses (1) the discovery and early work on the 
evolution of quasars and AGNs, (2) the different techniques used to find 
quasars and their suitability for evolutionary studies, (3) the current status 
of our knowledge of AGN evolution for $0<z<6$, (4) the new results and questions
that deep radio and X-ray surveys are producing for the subject, (5) the
relation of AGNs to the massive black holes being found in local galaxies
and what they tell us about both galaxy and AGN evolution, and (6) current
research problems and future directions in quasar and AGN evolution.
\end{abstract}

\section{Introduction and Background}

The subject of the evolution of the active galactic nucleus (AGN) population 
began in the late 1960s with Schmidt's (1968, 1970) discoveries that the space 
densities of both radio and optically selected quasars increased significantly with redshift.  The effect was 
so strong that it was detectable in samples as small as 20 objects.  He developed and 
applied the $V/V_m$ test for analyzing the space distribution in his samples and showed 
that there was a strong evolution in the space density of quasars toward higher redshift, 
increasing by more than a factor of 100 from redshift 0 to 2.  This was a striking 
and unexpected result that posed a question that is still crucial today --- What causes the 
sharp decline since $z=2$?

In this article I will review and discuss the following subjects:

\begin {itemize}

\item The techniques used to discover quasars and AGNs, their selection effects, and the 
surveys used to study the evolution of the AGN population.  
\item The general picture of evolution up to 1995, when the first well-defined, 
quantitative surveys of the evolution of high-redshift quasars were published.
\item The current status of major optical surveys such as 2dF and SDSS.
\item Radio and X-ray surveys and how they are critical to understanding AGN evolution.
\item The relation of AGNs to their host galaxies and how studies of massive black holes 
in spheroids provide constraints on AGN evolution.
\item Current research problems, such as measuring the quasar luminosity function at 
high redshift and faint magnitudes; relating observed to physical evolution; the 
framework for connecting observations, accretion processes, and the growth of black hole 
masses; and how to estimate black hole masses.

\end{itemize}

Before proceeding, let us define and discuss terms used in this article to aid the clarity of 
the presentation.  

An AGN is one not powered by normal stellar processes, 
although active star formation may be occurring in the vicinity.  The working hypothesis 
is that AGNs contain massive black holes and are powered by accretion processes.  Their 
luminosities range from as low as $M_B = -9$ mag  to as 
high as $M_B = -30$ mag ($L_X =10^{38}$ to $10^{48}$ erg s$^{-1}$).
Quasars are the high-luminosity ($M_B < -23$ mag, $L_X>10^{44}$ erg s$^{-1}$) 
members of the AGN family.

Traditionally, evolution of AGNs or quasars has meant the evolution with redshift of their 
luminosity function or space density (which is the integral of the luminosity function over 
some range of luminosities).  However, evolution can also refer to changes with redshift 
of the spectral energy distribution (SED) or the emission-line spectra of AGNs.  In 
general, {\it observed} evolution will refer to changes with redshift of any observed 
property of AGNs.

Ultimately, we wish to map and understand the {\it physical} evolution of AGNs, by 
which we mean how their central black holes form and grow with cosmic epoch and how 
their accretion processes and rates, which determine the luminosities and SEDs we 
observe from AGNs, evolve with cosmic epoch.  The discovery of the ubiquity of black 
holes in the spheroids of nearby galaxies makes us realize that the physical evolution of 
AGNs is closely connected with and is an important part of the larger subject of how 
galaxies in general form and evolve.  It appears that virtually every spheroidal system 
went through an AGN phase at some time in its history --- thus the subject of our meeting: 
``The Coevolution of Black Holes and Galaxies.''

However, the persistent question of how many AGNs are hidden because of weak 
emission lines, obscuration by dust, or absorption in X-rays has continued to impede 
progress in the mapping of the observational evolution of AGNs and must be addressed 
in any attempt to determine the properties of the overall AGN population.  Fortunately, 
the advent of powerful new space observatories such as {\it Chandra}\ and 
{\it XMM-Newton}, in 
conjunction with sensitive radio and infrared surveys, provides new tools for attacking 
this problem, as will be addressed below.  

At the same time, the formulation and application of the appropriate observational 
definitions of AGNs continue to be critical issues in current research, 
especially for low-luminosity objects, 
which can be hard to find within the glare of their host galaxy or to 
separate from normal stars.  For example, the work of Ho (2003) 
suggests that some AGNs may have X-ray luminosities down to $10^{36}$ erg s$^{-1}$, 
or less than 
stellar X-ray sources.  

If we are really to understand the global population of AGNs and their relation to 
galaxies, these problems must be solved.  This will be one of the themes to be developed 
in this article.

\section{Observational Techniques, Selection Effects, and Surveys}

The general principle for discovering AGNs is to make use of one or more of the ways in 
which they are not like stars or galaxies, for example, how they differ in the SEDs or
emission-line spectra.  
The pointlike, i.e., spatially unresolved, nature of the nuclei is 
another distinguishing factor.  It is also possible to make use of 
their great distances and 
correspondingly undetectable proper motions in quasar and AGN searches
or their variability in brightness.  In this article 
we will concentrate on techniques that make use of their SEDs and spectral-line 
properties.

The SEDs of AGNs are remarkable for their broad extent in frequency, 
from radio to $\gamma$-rays, 
which is much greater than for normal, thermal sources of astronomical radiation.  The 
UV/optical emission-line spectra stand out for the strength and breadth of the principal 
emission lines and for the wide range of ionization.  Typical line widths of permitted 
lines are 5000 km s$^{-1}$ or more.  The strongest individual lines are those of hydrogen 
(Ly$\alpha$, H$\alpha$, and H$\beta$), C~IV, C~III], Mg~II, and N~V, while broad emission 
complexes of Fe~II are visible.  In addition, forbidden lines of [O~I], [O~II], [O~III], and 
[S~II] are prominent.

Some of the observable properties of AGNs provide diagnostic probes of the physical 
nature of the central engine.  For example, the X-ray emission originates in regions as 
close as a few Schwarzschild radii of the central black hole and yields information about 
the inner part of the accretion disk and coronal region.  The broad UV/optical emission 
lines are produced within a few light days of the central engine.  One main goal of AGN 
research is to combine multiwavelength and spectral observations of AGNs with 
theoretical models of the accretion processes so that physical properties such as accretion 
rates and efficiencies can be inferred from observable data.  When success is achieved in 
this subject, it will yield a significant advance in our understanding of AGN evolution.

\subsection{Techniques for Finding Quasars and AGNs}
\subsubsection{Quasars}
Historically, quasars were first discovered (Hazard, Mackey,
\& Shimmins 1963; Schmidt 1963) via the optical identification of 
radio sources, a technique that 
was both effective and efficient because normal stars and galaxies are much weaker 
sources of radio emission.   
It was soon realized (Sandage 1965) that the bulk of the quasar population was radio 
quiet and could be identified through the excess UV (UVX) radiation 
that quasars demonstrated relative to normal stars\footnote{For the record, most stars are 
UV faint and quasars have relatively flat optical/UV SEDs in $\nu f_{\nu}$ space.}. 
We now know that the UVX technique is effective for redshifts up to about 2.2, 
the point at which Ly$\alpha$ emission shifts into the observed $B$ band and quasars 
begin to lose their characteristic UV excess.  We also know now
that only about 10\% of 
quasars and AGNs in the early samples were strong radio emitters, or radio-loud objects.

At higher redshifts, different techniques must be used to find quasars.  The problem 
becomes difficult for two reasons: (1) at redshifts around 3, the optical/UV SEDs of 
quasars are hard to distinguish from stars, and (2) the space density of quasars at $z>3$ 
declines rapidly with increasing redshift.  

The slitless-spectrum technique pioneered by Smith (1975) and developed by 
Osmer \& Smith (1976) provided a color-independent method of finding high-redshift 
quasars through the direct detection of their strong, broad emission lines, in 
particular Ly$\alpha$, on low-dispersion objective-prism photographs.  The technique 
was then applied to large telescopes through the use of a transmission grating/prism 
combination (grism) as the dispersing device (Hoag 1976).  
However, it was also realized that the 
slitless-spectrum technique was subject to an important selection effect in that it favored 
the detection of quasars with strong emission lines\footnote{Of course, all observational 
techniques for discovering quasars and AGNs are subject to selection effects.  This has 
been a long-standing problem in the determination of their luminosity function and its 
evolution.  Nonetheless, it appears that the slitless-spectrum technique indeed discovers 
the bulk of the high-$z$ population, although it obviously misses objects with weak or no 
emission lines.}.

Schmidt, Schneider, \& Gunn (1986) made an important advance on this 
problem by using a digital detector with a grism at the Hale
5-m telescope and by developing 
and applying a numerical selection algorithm for identifying emission-line objects whose 
properties and efficiencies could be quantified.  The effectiveness of their approach is 
well demonstrated in Figure 2 of their paper, 
which shows the grism spectra for a variety of high-redshift 
quasars from their survey.

The advent of rapid plate-scanning machines such as COSMOS and APM enabled the 
extension to higher redshift of color-based techniques for discovering quasars.  Warren et 
al. (1987) found the first quasar with $z=4$ in this way.  The machines and multi-color 
techniques made it possible to use more sophisticated combinations of colors to separate 
quasars from stars and to provide quantitative estimates of the selection efficiency as a 
function of redshift and apparent magnitude, which were crucial for determining the 
luminosity function.
Subsequently, the Sloan Digital Sky Survey (SDSS; York et al. 2000) combined the 
multicolor technique with a dedicated survey telescope and the largest digital camera 
built until that time to open a new frontier in extragalactic research by undertaking a 
digital survey of 10,000 deg$^2$ in five filters; the initial results are described in
more detail below.  

\subsubsection{AGNs}

We now know that the discovery of AGNs preceded quasars by 20 years 
(Seyfert 1943), although the connection and understanding was not achieved 
until the mid-1970s.  Seyfert's classic paper described the properties of nearby 
galaxies with unusually bright nuclei, which also had unusual emission-line spectra, in 
particular, broad lines and a wide range of ionization.  Seyfert galaxies and related AGNs 
such as LINERs (low-ionization nuclear emission-line regions), which are less 
luminous than $M_B = -23$ mag, constitute the bulk of the AGN population.  Their most 
prominent members can be discovered through imaging and spectroscopic surveys 
following in the footsteps of Seyfert.  However, the discovery of lower-luminosity, more 
elusive members of the class requires much more care, as the work of Ho, Filippenko, 
\& Sargent (1995) has shown.  They examined carefully {\it all} galaxies
within a magnitude-limited survey with high-
quality, narrow-slit spectra for evidence of an active nucleus.  

Most recently, Heckman (2003) and Hao \& Strauss (2003)
have  demonstrated that careful application of stellar population synthesis 
modeling to SDSS galaxy spectra can pull out otherwise unrecognizable emission-line 
and AGN signatures through the careful subtraction of the young stellar and nebular 
emission population.  Their work indicates that the presence of weak AGN 
activity is much more common than originally thought and is found in the
majority of early- and middle-type galaxies. %reaching possibly as high as 
%CHECK NUMBER \% of the total population.
%XXX is this true?  majority?

\subsubsection{Radio and X-Ray Techniques}

The discovery and identification of quasars and AGNs by radio and X-ray techniques is 
perhaps the most straightforward of all, because normal stars and galaxies are weak 
emitters in these wavelengths.  One requires sufficient sensitivity to compact sources and 
positional accuracies of $\sim 1^{\prime\prime}$ on the sky.  Objects in radio and/or X-ray 
catalogs are then matched to optical catalogs for identifications and follow-up optical 
spectra with a large telescope are used to confirm the identification and establish the 
redshift of the object.  Radio and hard X-ray sources offer the important advantage that 
they are not affected by dust obscuration that may occur along the line of sight to the 
AGN.  If spectral information is available in the 1 keV range, then estimates of the 
column density of any absorbing gas along the line of sight may be made.  

Until recently, radio surveys were hampered by the fact that, as mentioned previously, 
only about 10\% of AGNs are radio loud and thus radio surveys included only a small 
fraction of the total population.  However, with the advent of deep, wide-area surveys 
such as FIRST (Becker, White, \& Helfand 1995), 
important new opportunities have arisen.  FIRST, which 
reaches to milli-Jansky flux limits, is sufficiently sensitive to detect {\it radio-quiet} 
quasars.  When used in conjunction with the multi-color imaging data of the SDSS, it has 
enabled the discovery of new classes of AGNs (e.g., reddened broad absorption-line 
quasars that are radio sources) and added a new perspective on the issue of dust 
obscuration.  

The combination of FIRST and SDSS data overcomes another problem with earlier radio 
surveys, namely, the difficulty of achieving effective redshift preselection for candidate 
objects.  The difficulty was that follow-up spectroscopy of a large number of candidates 
had to be carried out to find high-redshift or rare types of quasars and AGNs.  However, 
the multicolor SDSS data now can be used to pre-sort candidate objects into the desired 
groups for follow-up work.  

X-ray data have been important to the study of quasars and AGNs since their first 
detections in X-rays because it was realized that the emission likely originated from very 
close to the central black hole.  Indeed, it can be argued that X-ray emission
is the defining characteristic of AGNs (e.g., Elvis et al. 1978).  
However, the point was somewhat moot 
at the time because of the lack of sensitivity of the original X-ray observatories.  
Now, following the work with {\it ROSAT}\ and the initial results from 
{\it Chandra}\ and
{\it XMM-Newton}, the 
tables are turned --- the deepest X-ray surveys are picking up objects not previously
noted in optical surveys. 

\subsubsection{Summary}

Discovery and survey techniques for quasars and AGNs at X-ray, 
UV/optical, and radio wavelengths are now sufficiently well developed, quantified, and 
sensitive that we have the main tools in hand to settle many of the most fundamental 
observational questions about the evolution of the AGN population.  The combinations of 
multiwavelength data that are now possible add even more opportunities for research on 
the nature of AGNs.

\section{Evolution of the AGN Population}

\subsection{Results through 1995}

Schmidt's discovery of the evolution of the quasar luminosity function immediately 
stimulated work on the nature of the evolution.  While powerful for showing the 
existence of evolution, the $V/V_m$ test by itself was not capable of delineating the 
nature of the evolution. Furthermore, the available quasar samples were too small to 
permit analyses in much detail.  Schmidt explored different forms of density evolution, 
i.e., evolution of the number density with cosmic epoch.  He found that both a power-law 
evolution of the form $(1+z)^k$ and an exponential function of look-back time could fit 
the data up to redshift 2.  Mathez (1976, 1978), building on the work of Lynds 
\& Petrosian 
(1972), demonstrated that luminosity evolution, in which the characteristic luminosity of 
quasars increased with redshift also provided a satisfactory fit to the data.  Schmidt
\& Green (1983) presented results from the 92 quasars in the Palomar Bright Quasar Survey
that showed the increase of space density with redshift to depend on the luminosity
of the objects.  This indicated that a simple parameterization of either pure density
or pure luminosity evolution did not fit the data well.

Subsequently, the work of Boyle, Shanks, \& Peterson (1988, hereafter BSP), using the UVX technique,
marked a significant advance in sample size and limiting magnitude.  
They compiled a sample of 420 quasars to $B<20.9$ mag
from UK Schmidt plates scanned with {\sc COSMOS}.  They found that a 
two-power law luminosity function and luminosity evolution adequately describe the data 
for objects with $M_B < -23$ mag and $z < 2.2$, a result that has been widely used and
is consistent with recent 2dF results, as described below.
 
Thus, the situation by the late 1980s was that either the space density of 
quasars increased by more than a factor of 100 between redshift 0 and 2 
(Schmidt \& Green 1983), or their 
characteristic luminosity increased by a factor of 30 (BSP). 

It was also understood that the density and luminosity evolution pictures led to 
significantly different estimates of the lifetimes of quasars, about $10^7$ years in the 
density evolution picture and $10^9$ years or more in the luminosity evolution picture.  
Another consequence was that most galaxies would pass through a quasar phase in the 
density evolution model; for luminosity evolution, only a few percent of galaxies would 
be active.

A different and also important question was raised early on in studies of quasar evolution: 
What happened at high redshift, $z>2$?  The redshift histograms of quasar catalogs 
showed a marked decline in numbers at $z>2$ (e.g., Hewitt \& Burbidge 1980), 
with the implication that the 
evolution also declined.  However, it was also realized that the traditional UVX method 
was not suitable for finding high-redshift quasars, and the lack of suitably defined 
samples blocked progress.  As mentioned above, the slitless-spectrum technique provided 
an efficient means of discovering high-redshift quasars, and Osmer (1982) showed from a 
differential study with the CTIO 4-m telescope and grism that there was strong evidence 
for a decline in the space density of quasars at $z>3$.  Nonetheless, he could only 
provide an upper limit on the decline because no quasars with $z>3$ were found in his 
survey, and it was clear that more work was needed.  Also, it was pointed out by Heisler 
\& Ostriker (1988) that dust absorption by intervening galaxies along 
the line of sight could produce a decline in the observed space density determined from 
flux-limited samples.

Significant advances occurred in the 1990s, when the first large, digital surveys for 
high-redshift quasars were carried out.  
Warren, Hewett, \& Osmer (1991a,b, 1994, hereafter WHO) 
made use of APM scans of UK Schmidt plates in six colors, $u, b_j, \upsilon, or, 
r, i$ to cover an effective area of 43 deg$^2$.  Their sample contained 86 objects with 
$16 < m_{or} < 20$ mag and $2.2 < z < 4.5$.  They developed a numerical modeling technique 
to determine the selection probabilities for their objects as a function of redshift and 
magnitude, allowing for different spectral slopes and emission-line strengths.  They used 
the selection probabilities to make two different estimates of the quasar luminosity 
function and its evolution.  They found strong evidence for a decline in the space density 
beyond $z=3.3$ by a factor of 6 for the interval $3.5 \leq z < 4.5$ for luminous quasars 
with $M_C < -25.6$ mag\footnote{$M_C$ is the absolute magnitude on the AB system for the
continuum level at Ly$\alpha$ (1216{\rm \AA}).}.

Schmidt, Schneider, \& Gunn (1995, hereafter SSG, and references therein) used the Palomar 
Transit Grism Survey (Schneider, Schmidt, \& Gunn 1994)
to establish a sample of 90 objects with Ly$\alpha$ emission and 
redshifts $2.75 < z <4.75$ to $AB_{1450} < 21.7$ mag in an area of 61.5 deg$^2$.  
Their digital survey used CCD 
detectors, and they determined the completeness and selection effects for their sample 
based on the line fluxes and signal-to-noise ratio of the data.  Their sample contained 8 
objects with $z>4$, and they found a decline in the space density of a factor of 2.7 per 
unit redshift for quasars with $M_B < -26$ mag and $z > 2.7$.
This result was very important because it used a survey technique
different from WHO and had many more quasars with $z>4$.

Kennefick, Djorgovski, \& de Carvalho (1995) made use of three colors, $J, F, N$, in the 
second Palomar Sky Survey in a program covering 681 deg$^2$ in the magnitude range 
$16.5 < r < 19.6$.  They had 10 quasars with $z>4$ in their sample and found a decline 
in space density of a factor of 7 at $z=4.35$ relative to $z=2.0$.

Taken together, the three surveys agreed well within their respective estimated errors and 
provided convincing evidence for a steep decline at $z>3$ in the observed space density 
of luminous, optically selected quasars.  
When the results are combined with those of BSP for lower 
redshifts and plotted on linear scales of space density versus look-back time 
%(Fig.~\ref{WHOSpDen}, {\it left}), the 
(Fig.~1.1, {\it left}), the 
behavior is dramatic and indicates a remarkable spike of quasar activity when the 
Universe was 15\%--20\% of its current age.

\begin{figure}
\centerline{
\hspace{-20mm}
\includegraphics[width=62mm,angle=0]{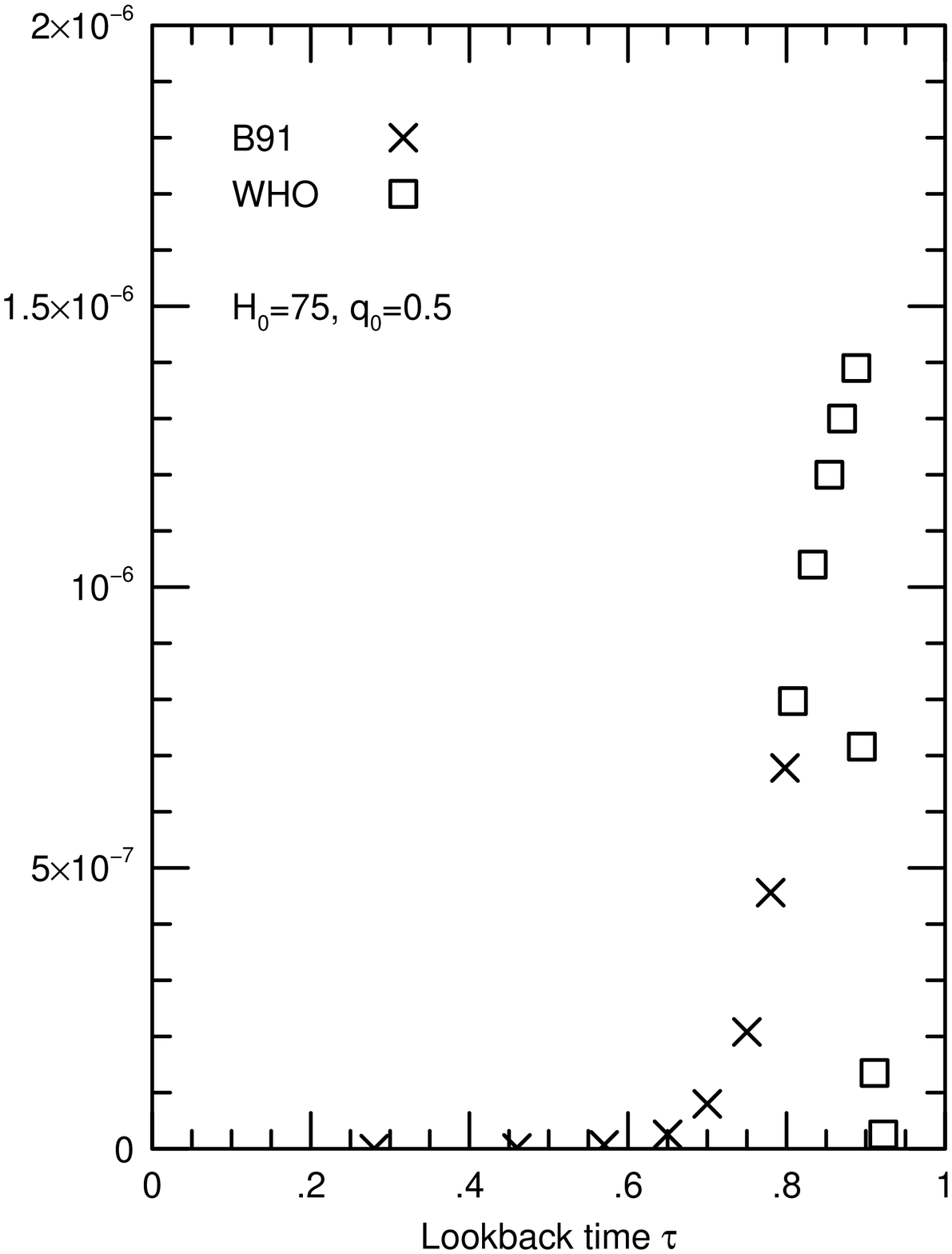}
\includegraphics[width=74mm,angle=0]{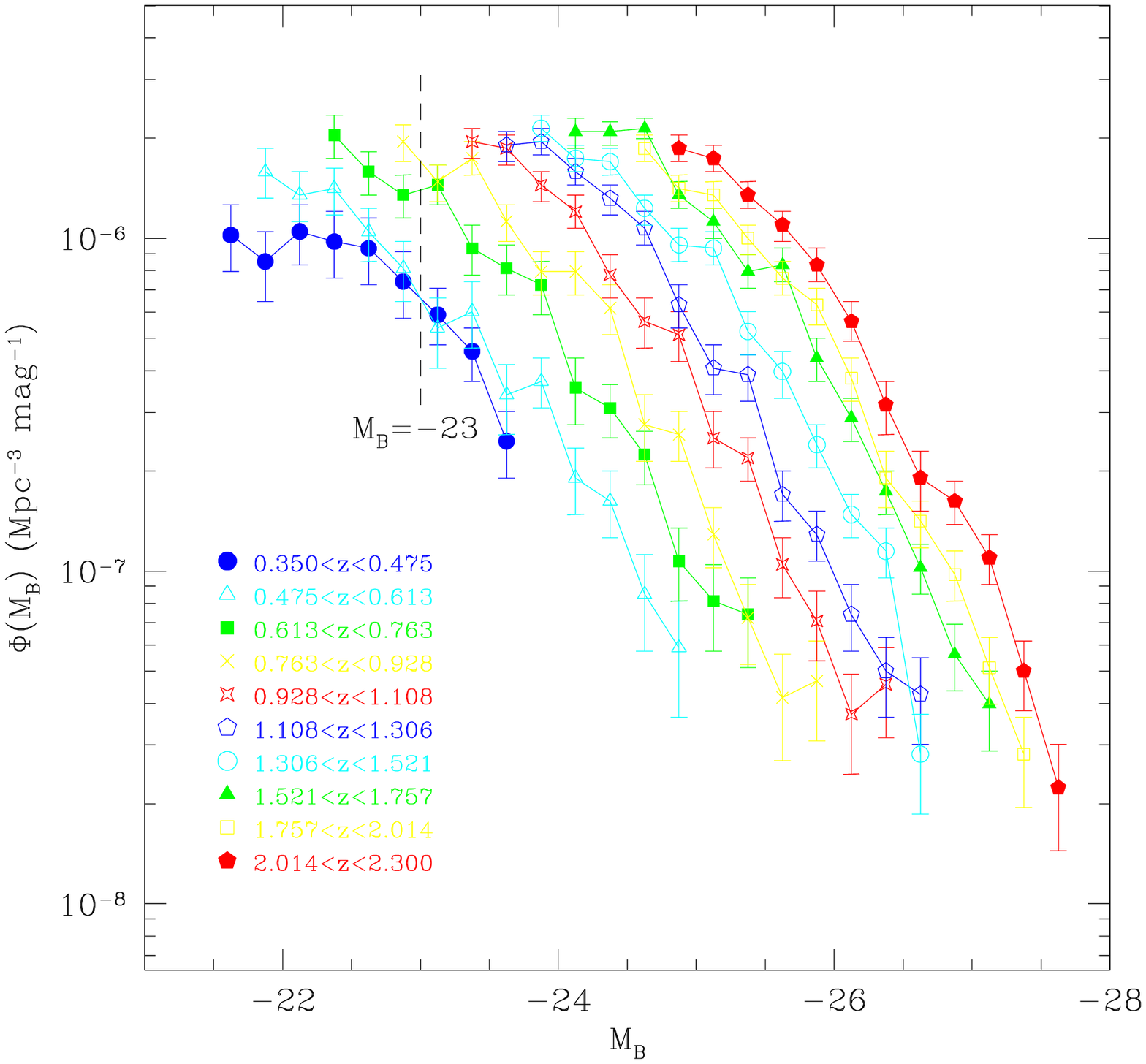}}
\caption{{\it Left:}\ A linear plot of the space density of luminous quasars
versus look-back time for the BSP and WHO samples. %\label{WHOSpDen}.
{\it Right:}\ The observed luminosity functions for the quasars
with $0.35 < z < 2.3$ in the 2dF sample compiled by Boyle
et al. (2000).} %\label{Boyle2000_4}.}
\end{figure}

At the same time, a number of important questions remained about the nature of the 
evolution of the quasar luminosity function: (1) At what redshift does the peak 
of the space density occur?  This is a result of the optical SEDs of $z=3$
quasars being similar to those of stars. (2) How do lower-luminosity quasars 
and AGNs, which
constitute the bulk of the population, evolve?  This requires deeper surveys.
(3) What is the form of the 
evolution and its possible dependence on redshift?  
Hewett, Chaffee, \& Foltz (1993) showed from a study of the 
1049 quasars and AGNs from the Large Bright Quasar Survey, which cover $0.2 < z < 3$ 
and $16.5 < m_{B_J} < 18.85$, that the data are not fit well by a pure luminosity evolution 
model with a two-power law luminosity function.  They found that the slope of the 
luminosity function became steeper at higher redshifts, the rate of evolution was slower 
for $0.2 < z <2$ than the Boyle et al. results, and the evolution continued, more slowly, 
until $z \approx 3$.  Thus, more work needs to be done.

\subsection{Recent Large Optical Surveys}

The above-mentioned questions on the evolution of quasars provided a significant
part of the motivation for two surveys significantly larger than anything 
previously attempted, the 2dF survey (Boyle et al. 2000, and references 
therein) and the SDSS (York et al. 2000).
The goal of the 2dF was to cover 750 deg$^2$ of sky to $B<21$ mag and find
$>25,000$ quasars with $z<2.3$ via the UVX technique.  The SDSS objective was to 
survey 10,000 deg$^2$ of sky in five colors, $u'g'r'i'z'$, to find 100,000 quasars
covering all redshifts up to 5.8.

Boyle et al. (2000) have estimated the luminosity
function from the first 6684 quasars in the 2dF quasar survey.
The survey was based on $u, b_J, r$ UK Schmidt plates and is
primarily a UVX technique.  It was estimated to be 90\% complete for $z<2$.  
Their final sample included 5057 objects from 196 deg$^2$ of sky
with $M_B<-23$ mag, $18.25<b_J<20.85$, 
$0.35<z<2.3$.  They combined their sample with 867 objects from
the LBQS (Hewett, Foltz, \&
Chaffee 1995) and fitted a two-power law form 
to the luminosity function data.   The data are shown in 
%Figure~\ref{Boyle2000_4}.
Figure~1.1 ({\it right}).
They found that a polynomial evolution 
of $L^{\ast}_{B(z)}$ fits the 
data well 
%(Fig.~\ref{Boyle2000_6}).  
(Fig.~1.2, {\it left}).
Thus, their main conclusion is that for 2dF+LBQS, $-26<M_B<-23\, {\rm mag}\,
(q_0=0.5)$ and $0.35<z<2.3$, pure luminosity evolution works fine, with the
characteristic luminosity of quasars increasing by a factor of 40 at the peak
of activity.  However,
the EQS (Edinburgh Quasar Survey, Miller et al., unpublished) 
and HEQS (Hamburg/ESO Quasar Survey, K\"ohler et al. 1997) do not fit as well, 
particularly for $z<0.5$, where the LBQS has few or no data.

\begin{figure}
\centerline{
\hspace{-17mm}
\includegraphics*[height=63mm,angle=0]{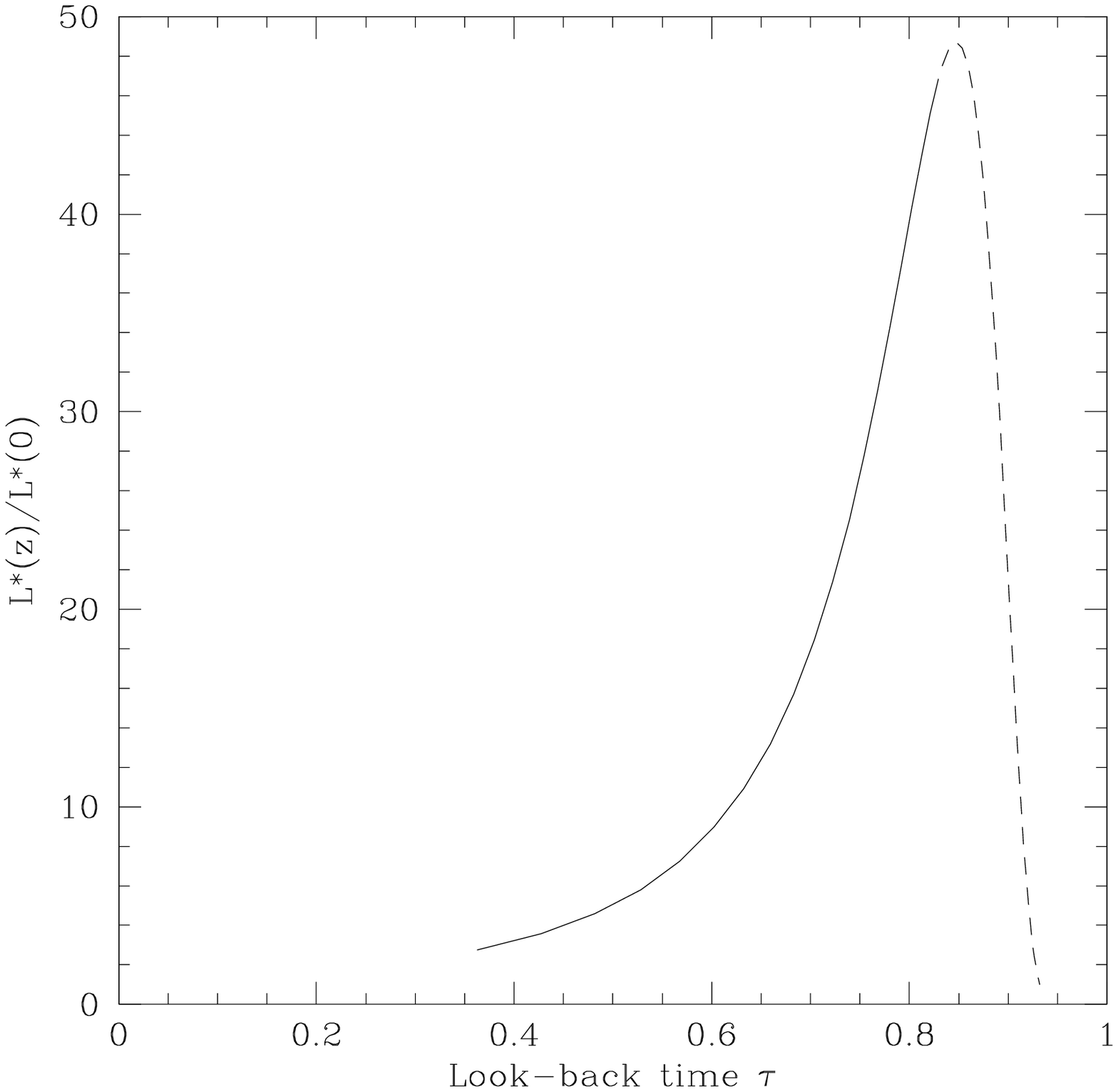}
\includegraphics*[width=68mm,angle=0]{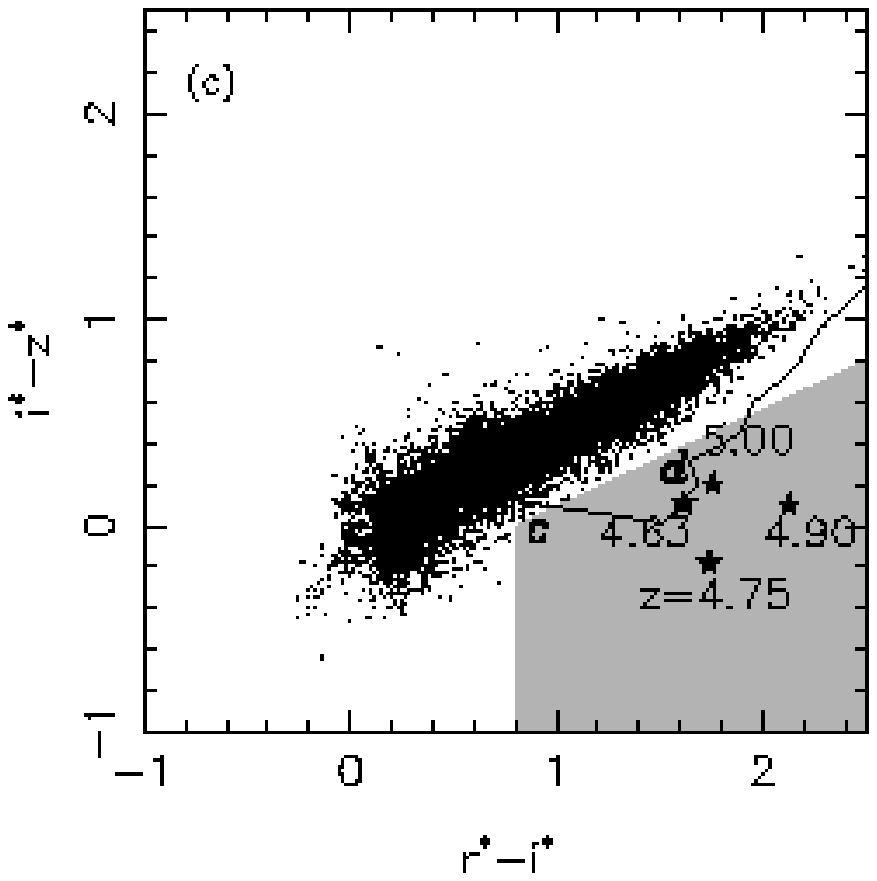}}

\caption{{\it Left:}\ A linear plot of the characteristic luminosity of
quasars from the 2dF (Boyle et al. 2000) and SSG samples
versus look-back time, for the luminosity evolution model.
%\label{Boyle2000_6}.
{\it Right:}\ A $i^{\ast}-z^{\ast}\, {\rm versus}\, r^{\ast}-i^{\ast}$ plot
of SDSS data, showing the stellar locus (black area and
points) and how quasars with $4.6<z<5.0$ separate from the
stellar locus because of the presence of Ly$\alpha$ emission
in the $i^{\ast}$ band and of Ly$\alpha$ forest absorption in
the $r^{\ast}$ band (Fan et al. 1999).} %\label{Fan1999}.}
\end{figure}

The SDSS opened an important window for the search for high-redshift quasars
through its large areal coverage of the sky and its use of the $z^{\ast}$ filter
with $\lambda_{{\rm eff}} \approx 9100${\rm \AA}, which extended the discovery space
in redshift to $z=5$ and beyond.  
%Figure~\ref{Fan1999} 
Figure~1.2 ({\it right})
(Fan et al. 1999) illustrates
how the $r^{\ast},i^{\ast},z^{\ast}$ filters were used to discover the first
quasars with $z \approx 5$ by their clear separation from the stellar locus.  

Fan et al. (2001a,b) then used the technique to compile a well-defined,
color-selected sample of 39 quasars with
$3.6<z<5.0$ and $i^{\ast} \leq 20$ mag in 182 deg$^2$ of sky.
They estimated the luminosity function for objects with $27.5 < M_{1450}<-25.5$
and its evolution with redshift.  
Their results, which are shown in 
%Figure~\ref{Fan2000_3}
Figure~1.3 ({\it left})
agree within the errors at $z \approx 4$ with the 
previous results of WHO, SSG, and KDC and give a value of the decline in space density 
of a factor of 3 per unit redshift for $z>3.6$.  They find a flatter slope for the 
luminosity function at $z \approx 4$ relative to earlier surveys for $z<3$.  These results 
confirm that pure luminosity evolution does not match the data between redshifts 2 and 5.

Then, Fan et al. (2001c) extended their work to discover four quasars 
with $z >5.8$ by using observation in the infrared J band to help eliminate nearby
and numerous L and T dwarf stars 
%(Fig.~\ref{Fan2001_2}).  
(Fig.~1.3, {\it right}).  
They showed that the space density at $z=6$ is about a factor of 
2 below that at $z=5$ and follows the decline with redshift just described.  
These objects provide the 
strongest evidence yet for a detection of the Gunn-Peterson absorption and thus evidence for
reionization at redshifts in the vicinity of $z \approx 6$\footnote{Note added in proof.  
Fan et al. (2003) announced the discovery of three additional quasars
with $z>6$.}.

\begin{figure}
\hspace{-5mm}
\centerline{
\includegraphics*[width=65mm]{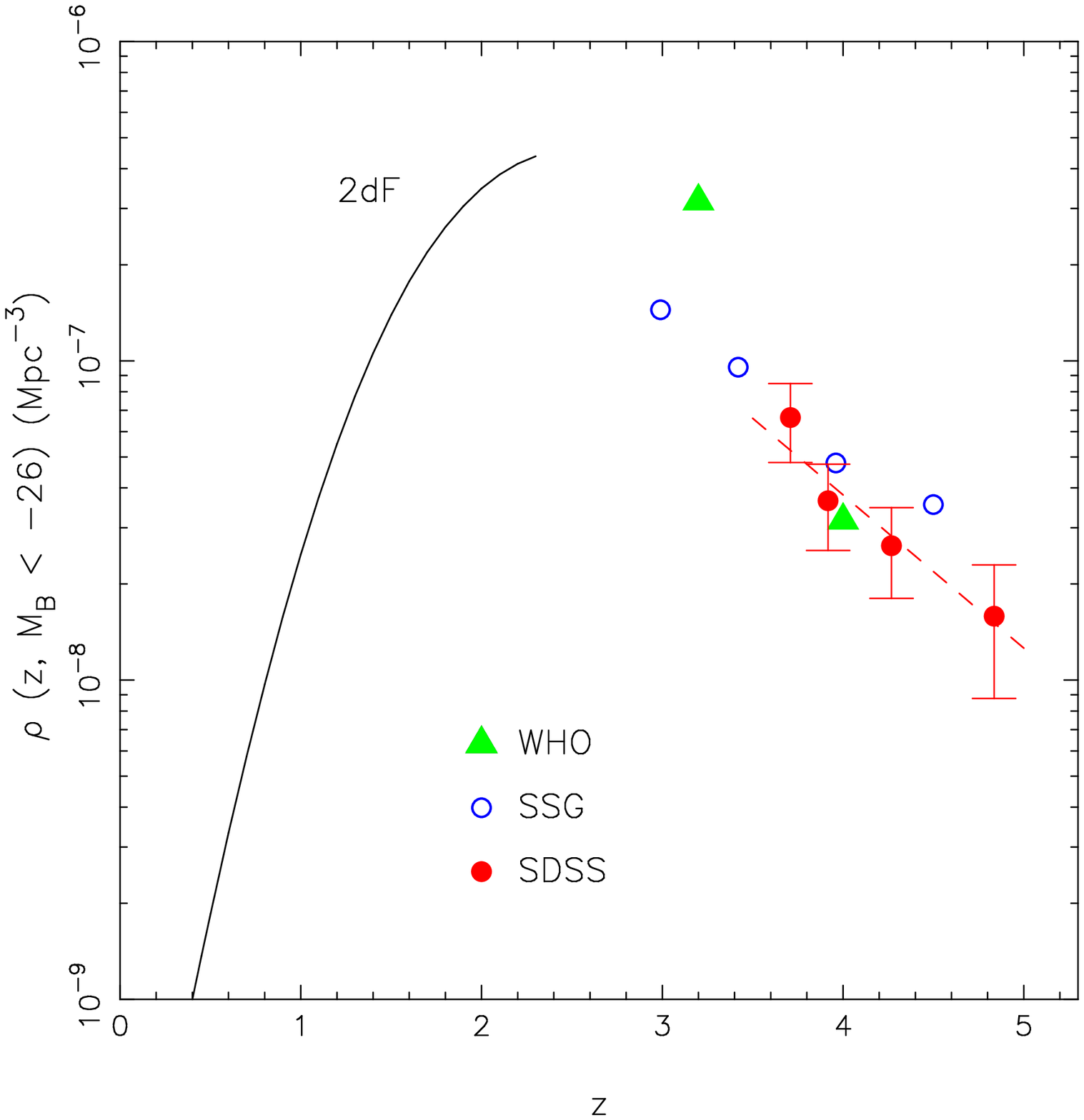}
\includegraphics[width=72mm,angle=0]{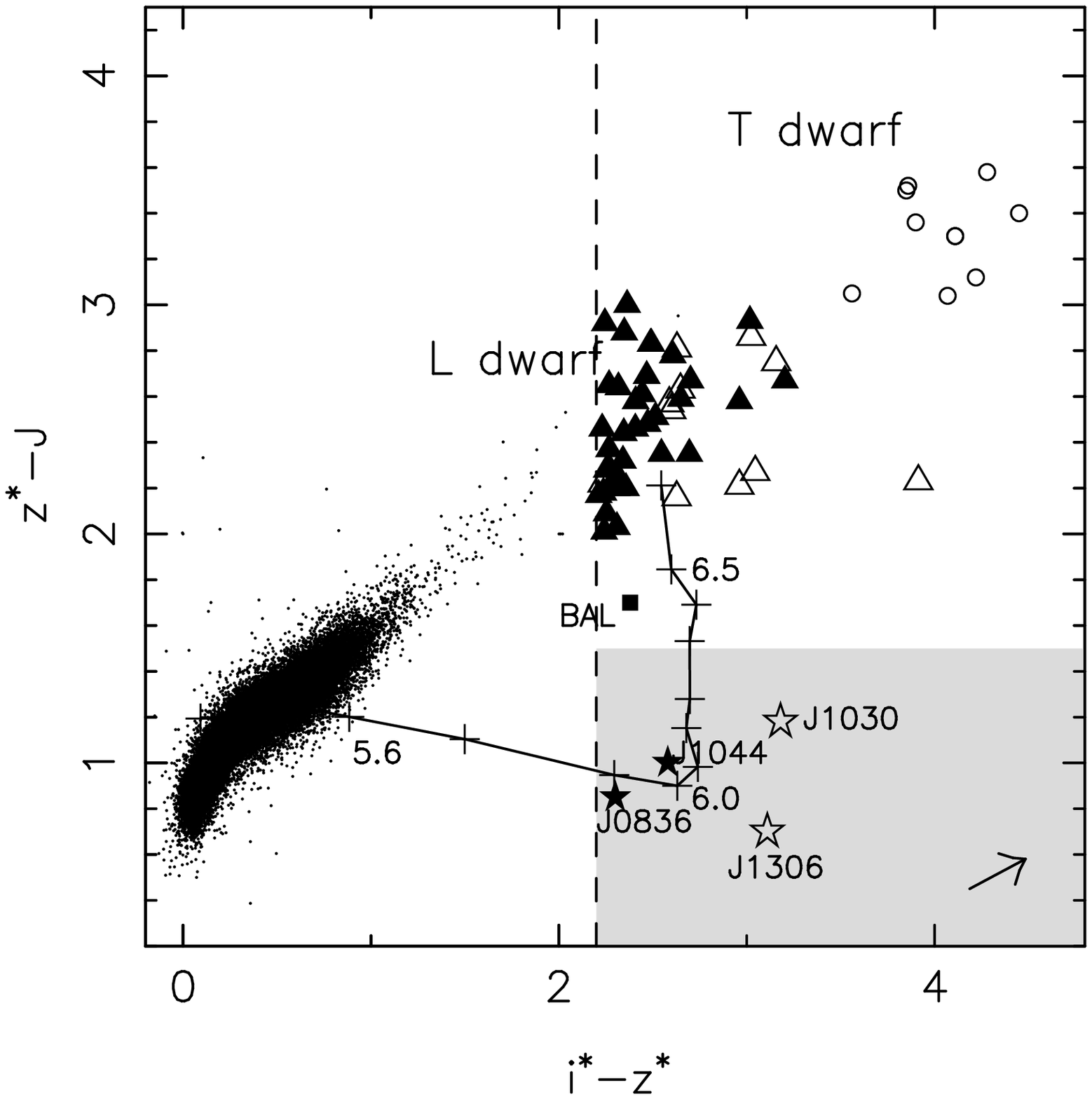}}
\caption{{\it Left:}\ The space density of quasars with $M_B<-26$ mag as a 
function of redshift for the 2dF, SDSS, SSG, and WHO surveys
(Fan et al. 2001b). %\label{Fan2000_3}.
{\it Right:}\ The $z^{\ast}-J \, {\rm versus}\, i^{\ast}-z^{\ast}$
diagram for the SDSS, showing how the infrared $J$ band enables
the separation of $z>5.8$ quasars from the numerous L and T dwarfs
(Fan et al. 2001c).} %\label{Fan2001_2}.}
\end{figure}

To summarize, the 2dF and SDSS results now cover the range $0.3 < z < 6.3$ 
for high-luminosity, optically selected quasars and 
reach close to the epoch of reionization. They provide by far the best data
on the evolution of such quasars that exist to date.

\subsection{Spectral Evolution}

We should also comment on the spectral properties of high-redshift quasars, which look 
surprisingly like their low-redshift counterparts.  
The emission-line spectra of the
first quasars discovered at $z \approx 5$ (e.g., Schneider, Schmidt, \& Gunn
1991; Fan et al. 1999) show C, N, O, and Si lines in the strengths
normally seen at low redshifts.
The Dietrich et al. (2002)
compilation of spectra covering $0.5<z<5$ shows remarkably little evolution 
with redshift.  While it is dangerous to jump to conclusions about abundances based on 
the appearance of strong emission lines, their lack of evolution is consistent with more 
detailed analyses.  Put another way, we have no evidence for chemical evolution in 
quasar spectra, except that, if anything, some abundances were {\it higher} at high 
redshifts (Hamann \& Ferland 1999).

\subsection{Evolution of Radio Sources}

At radio wavelengths, Hook, Shaver, \& McMahon (1998),
building on the work of Shaver et al. (1996),
have carried out an important survey that bears on the question of possible
obscuration by dust at high redshifts.  
Their sample contains 442 radio sources with $S_{\rm 2.7GHz} \geq 0.25$ Jy 
and stellar identifications.  The highest redshift object has $z=4.46$.  
For objects with radio power
$P_{\rm lim} > 7.2 \times 10^{26}$ W Hz$^{-1}$ sr$^{-1}$, they find an evolution of the space 
density very similar to WHO, SSG, and the SDSS.  
This is strong evidence against dust reddening 
being the main cause of the decline at high redshifts.

At the same time, Webster et al. (1995) and Gregg et al. (2002)
have argued that the finding of significant numbers of radio-selected
quasars with very red values of $B-K$ in the Parkes and FIRST
surveys indicates that up to 80\% of
the population is being missed in traditional optical surveys
because of dust obscuration.
This interpretation has been challenged by Benn et al. (1998)
and Whiting, Webster, \& Francis (2001) on the grounds that
the brightness in $K$  can arise from the emission of the host
galaxy and/or synchrotron radiation, not from obscuration
in the $B$ band by dust.  Until this issue can be resolved,
the question of dust-obscured quasars remains important to
our understanding of quasar evolution, as indicated by
the new X-ray results described below.  The finding of reddened
quasars in the 2MASS survey
(Marble et al. 2003, and references therein) is also contributing
important information on this subject.

\section{Estimating Black Hole Masses}

Until this point we have discussed the {\it observed} aspects
of the evolution of quasars and AGNs, primarily the
evolution of their luminosity functions.  Now let us begin
to consider their {\it physical} evolution, for which estimates of the
masses of the central black holes are crucial. Such estimates
will enable us to map the growth of the black holes with 
cosmic epoch.

In this meeting we have heard about three ways to estimate
black hole masses: (1) from their gravitational influence on the
stellar velocity distributions or gas kinematics
in the centers of galaxies, 
(2) reverberation mapping of the broad-line emission region
(Barth 2003), and (3) the use of emission-line
widths and continuum luminosities (e.g., C~IV, Vestergaard
2002, 2003; Mg~II, McClure \& Jarvis 2003).
The first two provide the underpinnings for the mass estimates
but are limited to nearby galaxies and AGNs.  The third method,
while indirect and subject to more uncertainties, has great
potential value because it provides the only practical way
we have at the moment of estimating the masses of quasars
and AGNs at high redshift.

Vestergaard \& Osmer (in preparation) are using methods 2 and 3
to make estimates of the mass functions of quasar samples
at low (the BQS, Schmidt \& Green 1983) and high
(SDSS, Fan et al. 2001a) redshift.  Their preliminary
results indicate that the SDSS quasars have already achieved
masses of $> 10^9\,M_\odot$ at $z>3.6$, and their cumulative mass density is
more than a order of magnitude above the BQS sample.  This
indicates that luminous quasars at high redshift built up their
masses early (see also Vestergaard 2003).  
The BQS cumulative mass density, on the other hand,
is an indicator of how the luminous activity has declined rapidly
by the present time, when luminous quasars are quite rare.
Interestingly, the SDSS cumulative mass densities appear to fit on
the extension of the results from  the Padovani, Burg, \& Edelson (1990) 
sample of Seyfert
galaxies at low redshift.  This is consistent with the idea
that both the low-redshift Seyferts and low-luminosity AGNs and 
the high-redshift SDSS quasars have achieved a substantial fraction
of their final black hole mass growth.

\section{Theoretical Considerations: How the Masses Grow}

All the recent observational data on quasars
and AGNs, in combination with theoretical studies of their
evolution and the accretion processes that produce both
their luminosity and growth in mass, are now enabling new
global studies of their history (e.g., Yu \& Tremaine 2002;
Yu 2003; Steed, Weinberg, \& Miralda-Escud\'{e}, in preparation).
The goal is to determine how an initial black hole mass
function evolves into the one observed today in the local Universe
by considering the continuity equation and how the the masses
grow with accretion processes.  The simple equation
$L=\epsilon (\dot{m}/\dot{m_{\rm Edd}}) M c^2$, where $L$ is the luminosity produced
by an accretion rate $\dot{m}$ in Eddington units
with efficiency $\epsilon$ for a black hole of mass $M$, tells
us that if we could observationally determine $L$ and $\epsilon$
along with black hole masses, for example, we would have enough 
information to model the evolution of the black holes in galaxies.
Put another way, the general goal is to combine the black hole
mass function, the time history of accretion, and the distribution of
accretion rates and efficiencies to see if we can match the 
observed luminosity and mass functions for AGNs and black holes.  One
immediate problem at present is that we do not have a way of
separately estimating $\epsilon$ and $\dot{m}/\dot{m_{\rm Edd}}$; typically people
assume that $\epsilon$ is 0.1 or some range of values depending
on the accretion models they adopt.  Another problem is accounting
properly for the number of obscured sources in flux-limited samples.

Nonetheless, there are enough existing data to permit interesting
progress on the problem.  For example,
the combination of the black hole mass function for local galaxies 
and the X-ray background provide integral constraints that must
be satisfied by any model.  The mass function represents the
end point of the accretion processes, while the X-ray background
provides a measure of the integrated luminosity produced by
accretion over the history of the Universe.  The improved optical
data on the quasar luminosity functions provide additional
constraints on how and when this all occurred, because they
map out the evolution of the emitted light with cosmic time.
At the same time, the deep X-ray and radio surveys and related optical
observations provide crucial information on the contribution
of obscured sources to the accretion history of the Universe.

Yu \& Tremaine (2002) find that the quasar luminosity functions
and local black hole mass functions are consistent if 
$\epsilon \approx 0.1$ and the black hole mass
growth occurred during the optically bright phase.  The lifetime
of luminous quasars would be of order $10^8$ years.  At the same
time, there remain important questions about the accretion efficiency
of lower luminosity quasars and AGNs and its dependence on accretion
rate, for example.

\section{Current Research Programs}

\subsection{Optical/Infrared Surveys}

Building on the success of 2dF and SDSS in delineating
the evolution of optically selected quasars at high luminosity,
a next logical and important observational step is to map the
evolution of lower luminosity objects.  They constitute the
bulk of the AGN population, and they are also crucial for
understanding the nature of the extragalactic ionizing background
radiation.  At high redshift, there is already evidence that the
numbers of quasars are too few to account for the observed
level of ionization of the intergalactic medium
(McDonald \& Miralda-Escud\'{e} 2001; Schirber \& Bullock 2003).

On the observational front, the slope of the luminosity function
of high-redshift quasars is quite uncertain.  Although Fan et al. (2001b)
have made estimates, and the upper limits on the number of
AGNs in the HDF also sets constraints (e.g., Conti et al. 1999), direct
observations are needed,
because it is the slope that determines the number
of faint AGNs.

Among the current surveys for fainter quasars at high redshift
are the BTC40 (Monier et al. 2002), 
COMBO-17 (Wolf et al. 2001), and the NOAO DEEP
survey (Januzzi \& Dey 1999).  All are
multi-color imaging surveys.  In addition, Steidel et al.
(2002) are investigating the AGN population found in a deep
spectroscopic survey of Lyman-break galaxies.  The main properties
of these surveys are:
\begin{enumerate}

\item BTC40 covered 40 deg$^2$ in $B, V, I$ and 36 deg$^2$ in $z$
and was designed to find quasars with $4.8<z<6$.  It reached
3$\sigma$ limiting magnitudes of $V=24.5$, $I=22.9$, and $z=22.9$.
To date it has yielded two quasars with redshifts of 4.6 and 4.8 and
produced candidates down to $I=22$ for future spectroscopy on 8--10 m
telescopes.

\item COMBO-17  uses 17 filters covering the 
$0.37 - 2.2\mu$ wavelength range to achieve in effect low-resolution
spectroscopy for a 1 deg$^2$ area down to $R=26$ mag.  This
survey, which will include 50,000 galaxies as well as
quasars and AGNs, is very ambitious and promises to
yield very significant results when completed.

\item The NOAO DEEP survey is covering 18 deg$^2$ of sky in $B_W,R,I,J,H,K$
to optical magnitudes of 26 and near-infrared magnitudes of 21.  Because
of its broad wavelength coverage and faint limiting magnitudes,
it will provide very important data on the evolution of AGNs.

\item The Steidel et al. survey complements the multi-color ones in that
it investigates the spectroscopic properties of galaxies at $z \approx 3$
down to $R_{AB} \approx 25.5$ mag.  It has found that about 3\% of the
galaxies are AGNs, many of which would not have been detected in deep
X-ray surveys, and thus is sampling a part of parameter space not covered
to date in other work.

\end{enumerate}

An additional value of all these surveys will be to combine their
results with those of the deep radio and X-ray surveys mentioned
above.  This will help define better the luminosity functions and
statistics of quasars, AGNs of different types, and normal
galaxies and thereby help improve our knowledge of
the evolution of all these objects.

\begin{figure}
\centerline{
\hspace{-10mm}
\includegraphics[height=120mm,angle=0]{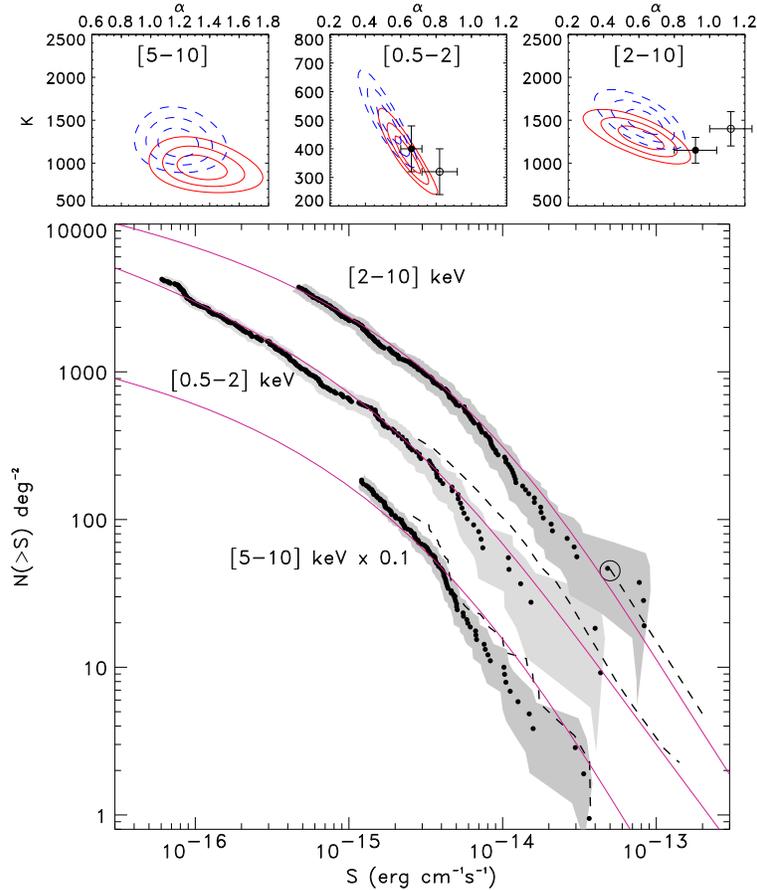}}
%\plotone{Fig7.ps}
\caption{The log $N$--log $S$ data for the {\it Chandra}\ Deep Field South
survey for three different energy bands.  The upper boxes show the
fits for the normalization constant, $\kappa$, and slope,
$\alpha$, of the faint end of the data (Rosati et al. 2002).
\label{Rosati2001}}
\end{figure}

\subsection{Evolution of X-ray Sources}

The {\it Chandra}\ and {\it XMM-Newton}\ observatories are enabling
X-ray surveys with sensitivity limits 2 orders of magnitude
fainter than was previously possible because of their 
large collecting areas and exposure times of a million seconds.  
In the {\it Chandra}\ Deep Field 
South (Rosati et al. 2002), 
%Figure~\ref{Rosati2001} 
Figure~1.4
shows that the 
surface density of sources is greater than 3000 deg$^{-2}$.
Thus, for the first time, the surface density of 
the deepest X-ray selected
AGNs exceeds the values of a few hundred deg$^{-2}$ that were
achieved in early deep optical surveys.
However, it is also now possible
to carry out optical imaging and spectroscopic observations
for sources that are 2 orders of magnitude fainter than
the nominal limit of SDSS, for example.  These capabilities
have led both to important discoveries and to the opening up of
important areas of research.  

For example, the deep source counts show that most of the
X-ray background (XRB) can be resolved and accounted for by faint
discrete sources.  However, to match the SED
of the XRB requires the existence
of a substantial number of absorbed AGNs at relatively low
redshifts (e.g., Gilli, Salvati, \& Hasinger 2001, building
on much previous work, such as that of Setti \& Woltjer 1989). 

Interestingly, optical identifications and follow-up spectroscopy
are now demonstrating the presence of these sources
(e.g., Barger et al. 2001; Hasinger private communication),
which have generally escaped notice in previous surveys either
because of their faintness or their unremarkable
optical appearance and spectra. Cowie et al.
(2003) and Hasinger (2003) have  assembled large enough samples
of objects to show clearly the excess of low-redshift AGNs
compared to the expectations of the evolutionary fits found
for the optical samples described above 
%(Fig.~\ref{HasingerFig6}).  
(Fig.~1.5).
Martini et al. (2002)
find an unexpectedly high fraction of X-ray selected AGNs
in the cluster Abell 2104, only one of which has the characteristic
emission lines of an AGN in its optical spectrum. These studies
demonstrate the important power that deep X-ray observations bring
to the studies of low-luminosity AGNs.  

\begin{figure}
\centerline{
\hspace{-20mm}
\includegraphics*[height=43mm,angle=0]{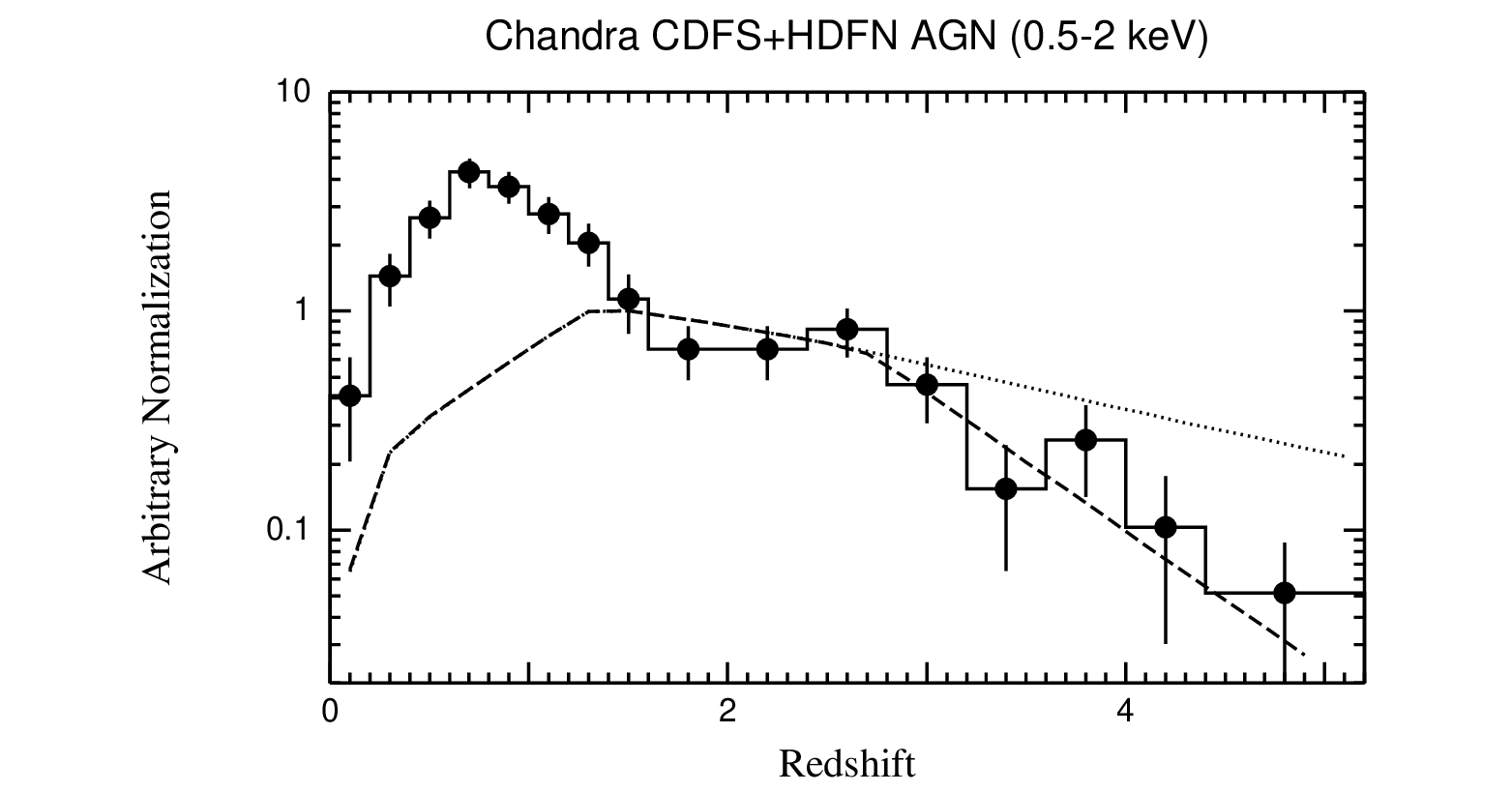}
\includegraphics*[height=42mm,angle=0]{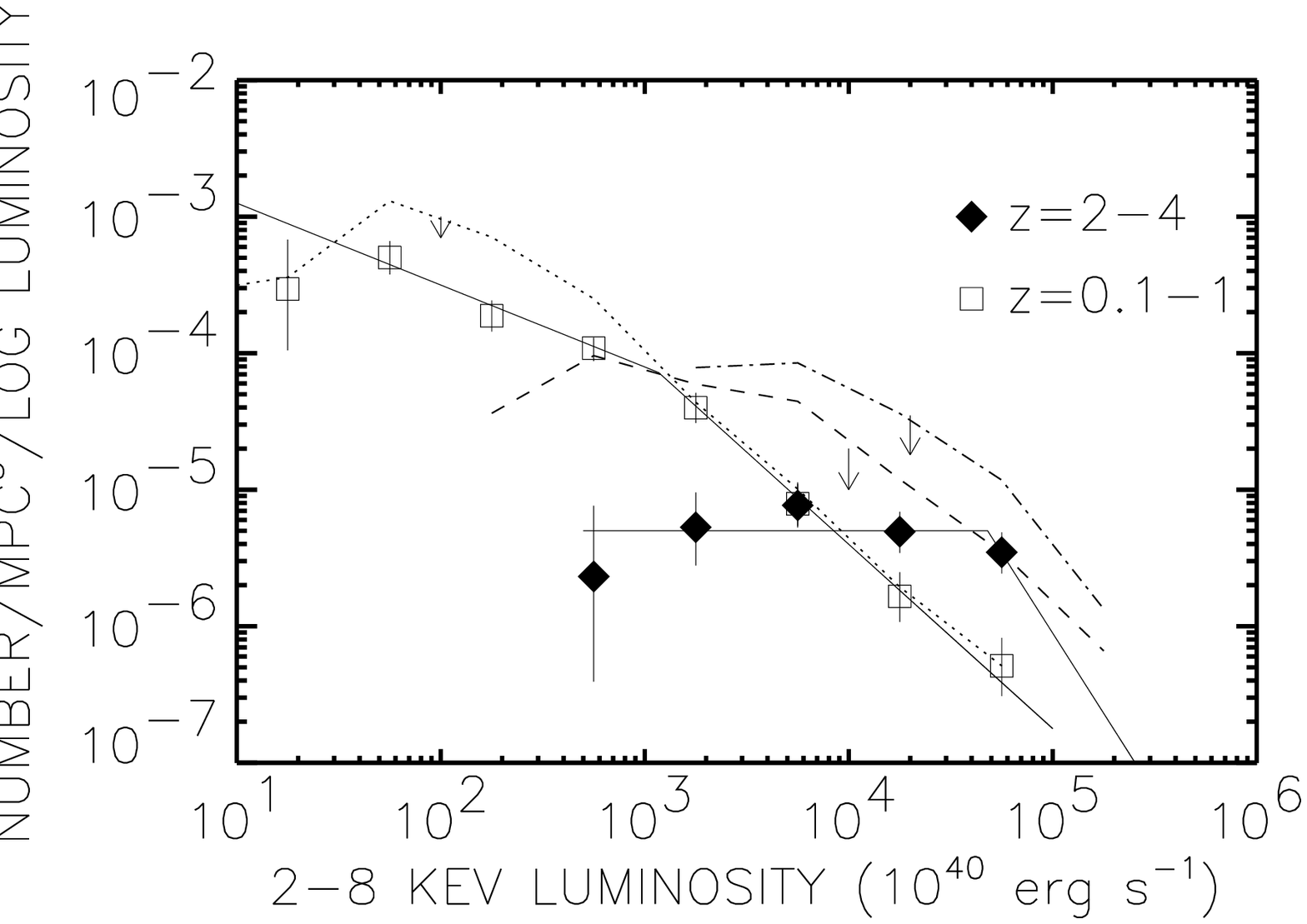}
}
\noindent
\caption{{\it Left:} The redshift distribution for 243 AGNs in the {\it Chandra}\
Deep Field South and Hubble Deep Field North surveys, from Hasinger (2003), compared to 
population synthesis models by Gilli et al. (2001), where the
dashed line is for the redshift decline of SSG for $z>2.7$ and 
the dotted line is for a constant space density for $z>1.5$.
Note the observed excess of objects at $z<1$.
{\it Right:} Luminosity functions derived from 
{\it Chandra}, {\it ROSAT}, and {\it ASCA}\
surveys by Cowie et al. (2003), based on their redshift measurements
and estimates.  Note that the values for sources with $z<1$ and 
$L_X<10^{43}$ erg s$^{-1}$ are well above those for the
$2<z<4$ objects.\label{HasingerFig6}}
\end{figure}

Thus, we are making good progress on mapping both the total contribution
of AGNs to the X-ray background and the evolution with time of their
X-ray emission, which in turn will lead to a measure of the accretion history
of discrete sources in the Universe.   We may look forward to substantial
advances in the next decade as the observational data continue
to improve and the physics of accretion is better understood.

\section{Next Steps}

In looking ahead, we can see that important next steps
in this field include:

\begin{itemize}

\item Completing the mapping of the X-ray and optical luminosity
functions for quasars and AGNs down to luminosities that include
the bulk of the integrated radiation.

\item Refining methods for mass determinations and applying them to the
full observed range of redshifts and luminosities.

\item Finding an observable spectral signature for accretion modes and
efficiencies that will allow us to make reliable estimates
of accretion rates.
Are low-luminosity AGNs a result of low accretion rate, low
efficiency, or low mass?

\item Determining the numbers of obscured sources and establishing
the correlation between, for example, absorption in X-rays 
and UV/optical obscuration by dust.  See if the results
are consistent with deep radio and sub-mm observations.

\item Achieving a self-consistent fit of the population of observed,
discrete X-ray sources with the overall intensity level and SED of
the X-ray background.

\end{itemize}

If, in the end, we can match the observational data for AGNs
over their entire redshift range to the local mass function
of black holes in galaxies, we will have made a significant leap
in our understanding of the coevolution of black holes and galaxies.

\vspace{0.3cm}
{\bf Acknowledgements}. I am grateful to Eric Monier for assistance with
the preparation of this article, especially the figures, and to David
Weinberg, Marianne Vestergaard, Brad Peterson, and the anonymous
referee for valuable comments on the first drafts.  I thank the organizers
for the opportunity to speak at the meeting.


\begin{thereferences}{}

\bibitem{}
Barger, A. J., Cowie, L. L., Bautz, M. W., Brandt, W. N., Garmire, G. P.,
Hornschemeier, A. E., Ivison, R. J., \& Owen, F. N. 2001, \aj, 122, 2177

\bibitem{}
Barth, A.~J. 2003, in Carnegie Observatories Astrophysics Series, Vol. 1:
Coevolution of Black Holes and Galaxies, ed. L. C. Ho (Cambridge:
Cambridge Univ. Press)

\bibitem{}
Becker, R.~H, White, R.~L., \& Helfand, D.~J. 1995, \apj, 450, 559

\bibitem{}
Benn, C.~R., Vigotti, M., Carballo, R., Gonzalez-Serrano, J.~I., \&
S\'anchez, S.~F. 1998, \mnras, 495, 451
 
\bibitem{}
Boyle, B. J., Shanks, T., Croom, S. M., Smith, R. J., Miller, L., Loaring, N.,
\& Heymans, C. 2000, \mnras, 317, 1014

\bibitem{}
Boyle, B. J., Shanks, T., \& Peterson, B. A. 1988, \mnras, 235, 935

\bibitem{}
Conti, A., Kennefick, J. D., Martini, P., \& Osmer, P. S. 1999, \aj, 117, 645

\bibitem{}
Cowie, L. L., Barger, A. J., Bautz, M. W., Brandt, W. N., \& Garmire, G. P. 
2003, \apj, 584, L57

\bibitem{}
Dietrich, M., Hamann, F., Shields, J. C., Constantin, A., Vestergaard, M., 
Chaffee, F., Foltz, C. B., \& Junkkarinen, V. T. 2002, \apj, 581, 912

\bibitem{}
Elvis, M., Maccacaro, T., Wilson, A. S., Ward, M. J., Penston, M. V.,
\& Fosbury, R. A. E. 1978, \mnras, 183, 129

\bibitem{}
Fan, X. et al. 1999, \aj, 118, 1

\bibitem{}
------. 2001a, \aj, 121, 31

\bibitem{}
------. 2001b, \aj, 121, 54

\bibitem{}
------. 2001c, \aj, 122, 2833

\bibitem{}
------. 2003, \aj, 125, 1649

\bibitem{}
Gregg, M.~D., Lacy, M., White, R.~L., Glikman, E., Helfand, D.~J., Becker,
R.~H., \& Brotherton, M.~S. 2002, \apj, 564, 133

\bibitem{}
Gilli, R., Salvati, M., \& Hasinger, G. 2001, \aa, 366, 407

\bibitem{}
Hamann, F., \& Ferland, G. 1999, \annrev, 37, 487

\bibitem{}
Hao, L., \& Strauss, M.~A. 2003, in Carnegie Observatories Astrophysics
Series, Vol. 1: Coevolution of Black Holes and Galaxies, ed. L. C. Ho
(Pasadena: Carnegie Observatories,
http://www.ociw.edu/ociw/symposia/series/symposium1/proceedings.html)

\bibitem{}
Hasinger, G. 2003, in IAU SYmp. 214, High Energy Processes and Phenomena in 
Astrophysics, ed.  X. Li, Z. Wang, \& V. Trimble (San Francisco: ASP), in press
(astro-ph/0301040)

\bibitem{}
Hazard, C., Mackey, M. B., \& Shimmins, A. J. 1963, \nat, 197, 1037

\bibitem{}
Heckman, T. 2003, in Carnegie Observatories Astrophysics Series, Vol. 1:
Coevolution of Black Holes and Galaxies, ed. L. C. Ho (Cambridge:
Cambridge Univ. Press)

\bibitem{}
Heisler, J., \& Ostriker, J. P. 1988, \apj, 332, 543

\bibitem{}
Hewett, P. C., Foltz, C. B., \& Chaffee, F. H. 1993, \apj, 406, L43

\bibitem{}
------. 1995, \aj, 109, 1498

\bibitem{}
Hewitt, A., \& Burbidge, G. 1980, \apjs, 43, 57

\bibitem{}
Ho, L.~C. 2003, in IAU Colloq. 184, AGN Surveys, ed. R.~F. Green, E.~Ye.
Khachikian, \& D.~B. Sanders (San Francisco: ASP), 13

\bibitem{}
Ho, L. C., Filippenko, A. V., \& Sargent, W. L. W. 1995, \apjs, 98, 477

\bibitem{}
Hoag, A. A. 1976, \pasp, 88, 860

\bibitem{}
Hook, I. M., Shaver, P., \& McMahon, R. G. 1998, in The Young Universe: 
Galaxy Formation and Evolution at Intermediate and High Redshift,
ed. S. D'Odorico, A. Fontana, \& E. Giallongo (San Francisco: ASP), 17

\bibitem{}
Jannuzi, B. \& Dey, A. 1999, in The Hy-Redshift Universe: Galaxy Formation and 
Evolution at High Redshift, ed. A. J. Bunker \& W. J. M. van Breugel 
(San Francisco: ASP), 258

\bibitem{}
Kennefick, J. D., Djorgovski, S. G., \& de Carvalho, R. R. 1995, \aj, 110, 2553

\bibitem{}
K\"ohler, T., Groote, D., Reimers, D., \& Wisotzki, L.  1997, \aa, 325, 502

\bibitem{}
Lynds, R., \& Petrosian, V. 1972, \apj, 175, 591

\bibitem{}
Marble, A.~R., Hines, D.~C.,  Schmidt, G. D., Smith, P. S., Surace, J. A.,
Armus, L., Cutri, R. C., \& Nelson, B. O. 2003, \apj, in press

\bibitem{}
Martini, P., Kelson, D. D., Mulchaey, J. S., \& Trager, S. C.
2002, \apj, 576, L109

\bibitem{}
Mathez, G. 1976, \aa, 53, 15

\bibitem{}
------. 1978, \aa, 68, 17

\bibitem{}
McClure, R.~J., \& Jarvis, M. J. 2003, in Carnegie Observatories Astrophysics 
Series, Vol. 1: Coevolution of Black Holes and Galaxies, ed. L. C. Ho 
(Pasadena: Carnegie Observatories, 
http://www.ociw.edu/symposia/series/symposuium1/proceedings.html)

\bibitem{}
McDonald, P., \& Miralda-Escud\'{e}, J. 2001, \apj, 549, L11

\bibitem{}
Monier, E. M., Kennefick, J. D., Hall, P. B., Osmer, P. S.,
Smith, M. G., Dalton, G. B., \& Green, R. F. 2002, \aj, 124, 2971

\bibitem{}
Osmer, P. S. 1982, \apj, 253, 28

\bibitem{}
Osmer, P. S., \& Smith, M. G. 1976, \apj, 210, 267

\bibitem{}
Padovani, P., Burg. R., \& Edelson, R. A. 1990, \apj, 353, 438

\bibitem{}
Rosati, P., et al. 2002, \apj, 566, 667

\bibitem{}
Sandage, A. 1965, \apj, 141, 328

\bibitem{}
Schirber, M., \& Bullock, J. S. 2003, \apj, 584, 110

\bibitem{}
Schmidt, M. 1963, \nat, 197, 1040

\bibitem{}
------. 1968, \apj, 151, 393

\bibitem{}
------. 1970, \apj, 162, 371

\bibitem{}
Schmidt, M., \& Green, R. F. 1983, \apj, 269, 352

\bibitem{}
Schmidt, M., Schneider, D. P., \& Gunn, J. E. 1986, \apj, 306, 411

\bibitem{}
 ------. 1995. \aj, 110, 68

\bibitem{}
Schneider, D. P., Schmidt, M., \& Gunn, J. E. 1991, \aj, 102, 837 

\bibitem{}
------. 1994, \aj, 107, 1245

\bibitem{}
Setti, G., \& Woltjer, L. 1989, \aa, 224, L21

\bibitem{}
Seyfert, C.~K. 1943, \apj, 97, 28

\bibitem{}
Shaver, P. A., Wall, J. V., Kellermann, K. I., Jackson, C. A.,
\& Hawkins, M. R. S. 1996, \nat, 384, 439

\bibitem{}
Smith, M. 1975, \apj, 202, 591

\bibitem{}
Steidel, C. C., Hunt, M. P., Shapley, A. E., Adelberger, K. L.,
Pettini, M., Dickinson, M., \& Giavalisco, M. 2002, \apj, 576, 653

\bibitem{}
Vestergaard M., 2002, \apj, 571, 733

\bibitem{}
------. 2003, in Carnegie Observatories Astrophysics Series, Vol. 1:
Coevolution of Black Holes and Galaxies, ed. L. C. Ho (Pasadena: Carnegie
Observatories, http://www.ociw.edu/symposia/series/symposuium1/proceedings.html)

\bibitem{}
Warren, S. J., Hewett, P. C., Irwin, M. J., McMahon, R. G.,
\& Bridgeland, M. T. 1987, \nat, 325, 131

\bibitem{}
Warren, S. J., Hewett, P. C., \& Osmer, P. S. 1991a, \apjs, 76, 1

\bibitem{}
------. 1991b, \apjs, 76, 23

\bibitem{}
------. 1994, \apj, 421, 412

\bibitem{}
Webster, R. L., Francis, P. J., Peterson, B. A., Drinkwater, M. J.,
\& Masci, F. J. 1995, \nat, 375, 469

\bibitem{}
Whiting, M.~T., Webster, R.~L., \& Francis, P.~J. 2001, \mnras, 323, 718

\bibitem{}
Wolf, C., Borch, A., Meisenheimer, K., Rix, H.-W., Kleinheinrich, M., \& Dye, 
S. 2001, Astronomische Gesellschaft Abstract Series, Vol. 18., 
abstract MS 05 39 

\bibitem{}
York, D., et al. 2000, \aj, 120, 1579

\bibitem{}
Yu, Q. 2003, in Carnegie Observatories Astrophysics Series, Vol. 1:
Coevolution of Black Holes and Galaxies, ed. L. C. Ho (Pasadena: Carnegie
Observatories, http://www.ociw.edu/symposia/series/symposuium1/proceedings.html)

\bibitem{}
Yu, Q., \& Tremaine, S. 2002, \mnras, 335, 965

\end{thereferences}
\end{document}